\documentclass{IEEEtran}
\usepackage{cite}
\usepackage{amsmath,amssymb,amsfonts}
\usepackage{algorithmic}
\usepackage{mathtools}
\usepackage{graphicx}
\usepackage{textcomp}
\usepackage{gensymb}
\usepackage{xcolor}
\usepackage[font=small]{subcaption}
\usepackage[font=small]{caption}
\usepackage{multicol}
\def\BibTeX{{\rm B\kern-.05em{\sc i\kern-.025em b}\kern-.08em
    T\kern-.1667em\lower.7ex\hbox{E}\kern-.125emX}}
    
\begin{document}

\title{Design of Compact Huygens' Metasurface Pairs with Multiple Reflections for Arbitrary Wave Transformations }
\author{Vasileios G. Ataloglou, \IEEEmembership{Student Member, IEEE}, Ayman H. Dorrah, \IEEEmembership{Student Member, IEEE}, and George~V.~Eleftheriades, \IEEEmembership{Fellow, IEEE}
\thanks{The authors are with The Edward S. Rogers Sr. Department of Electrical and
Computer Engineering, University of Toronto, Toronto, ON M5S 3G4, Canada
(e-mail: vasilis.ataloglou@mail.utoronto.ca; ayman.dorrah@mail.utoronto.ca; gelefth@waves.utoronto.ca).}
}
\pagenumbering{arabic}
\maketitle

\begin{abstract}
Huygens’ metasurfaces have demonstrated a remarkable potential to perform wave transformations within a subwavelength region. In particular, omega-bianisotropic Huygens’ metasurfaces have allowed for the passive implementation of any wave transformation that conserves real power locally. Previous reports have also shown that Huygens’ metasurface pairs are capable of realizing transformations that break the local power conservation requirement by redistributing the total power, while the wave propagates between the two metasurfaces. However, the required separation distance overshadows the low-profile characteristics of the individual metasurfaces and leads to bulky designs, especially for lower frequencies. In this work, we develop a method of designing omega-bianisotropic Huygens’ metasurface pairs, relying on a point-matching process of the real power at the two metasurfaces. We highlight the versatility of our method by presenting two variations of the configuration, depending on whether the electromagnetic source is located within or outside the metasurface pair. Based on the
examples of a cylindrical-wave to plane-wave transformation and a beam expander, we examine the impact of multiple reflections, as a way to overcome the size limitations and design compact structures. Moreover, we explore possible beamforming applications through an example of a Taylor-pattern antenna with a single feed-point between the two metasurfaces.
\end{abstract}


\section{Introduction}
Huygens' metasurfaces are electrically thin devices that have attracted considerable attention as an efficient tool to manipulate electromagnetic waves at will \cite{Holloway:TAP2012,Pfeiffer:PRL2013,Selvanayagam}. In their passive form, they consist of sub-wavelength elements (unit cells) arranged in a thin sheet, which induce equivalent electric and magnetic currents, when excited by an incident wave. The current densities that support the desired electric and magnetic fields at the two sides of the metasurface are calculated using the generalized sheet transition conditions (GSTC) \cite{Kuester:TAP2003}. Subsequently, the current densities are discretized across the metasurface and the response of each individual unit cell is engineered in terms of polarizabilities, susceptibilities or surface impedances/admittances \cite{Kuester:TAP2003, Achouri:TAP2015, Epstein:JOSAB2016}. This approach has led to the design and experimental demonstration of passive Huygens' metasurfaces (HMS) for numerous electromagnetic applications, such as engineering refraction, reflection and absorption, beam focusing and polarization control among others \cite{Pfeiffer:PRL2013,Selvanayagam,JPSWong:Nano2014,Estakhri:PRX2016,Radi:PRApplied2015,Chen:TAP2019Lens,Niemi:TAP2013,Pfeiffer:PRA2014}.

In the effort to realize more complicated wave transformations, omega-bianisotropy has emerged as a way to introduce another degree of freedom in the design process, accounting for magnetoelectric coupling \cite{Epstein:TAP2016,Asadchy:PRB2016,Epstein:TAP2017Antennas}. Specifically, in omega-bianisotropic Huygens' metasurfaces, the electric and magnetic fields both excite orthogonally-polarized equivalent electric and magnetic currents. Under this condition, it has been theoretically proven that a passive and lossless HMS can be designed for any transformation, as long as the real power density propagating normal to the boundary is locally conserved \cite{Epstein:TAP2016}. Previously impossible to realize wave transformations, such as perfect wide-angle reflectionless refraction, were successfully demonstrated with the use of omega-bianisotropic Huygens' metasurfaces \cite{Lavigne:TAP2018, Chen:PRB2018, Chen:TAP2019WideAngle}.

Despite the extra degree of freedom provided by the omega-bianisotropy, the condition of local power conservation along the metasurface places constraints on the form of the input and output electric and magnetic fields. Several solutions have been proposed as a way to overcome these limitations and achieve nonlocal power conserving transformations. For instance, surface waves have been used either as auxiliary fields to restore power matching locally without interfering with the far-field characteristics of the transformation or as a way to redistribute the power within metasurface systems that are based on the conversion of propagating waves to surface waves and vice versa \cite{Epstein:PRL2016,Kwon:PRB2018,Achouri:SCI2018}. However, the use of auxiliary surface waves requires specifying their form, which can be a non-trivial task for some applications (e.g. beamforming)\cite{Epstein:PRL2016}. On the other hand, metasurface systems usually suffer from low conversion efficiency between surface and propagating waves, as well as from losses and distortion at the discontinuities between the different regions of the system\cite{Achouri:SCI2018}. 

Recently, two cascaded omega-bianisotropic Huygens' metasurfaces have been studied for performing wave transformations that break the local power conservation condition\cite{Raeker:PRL2019, Ayman:AWPL2018}. As the wave propagates between the two metasurfaces, the power density profile is reshaped. Therefore, it is possible to satisfy local power conservation for the two metasurfaces individually, even if the input and output power density profiles of the total structure are different. One major disadvantage of this approach is the size requirements for the metasurface pair, as a substantial separation may be required, especially for the cases that the input and output power density profiles differ significantly. Although in \cite{Ayman:AWPL2018} it was stated that the introduction of multiple reflections may reduce the size requirements, their effect was not investigated.

In this paper, we design pairs of omega-bianisotropic HMS for arbitrary wave transformations that do not conserve local power. As in \cite{Ayman:AWPL2018}, the design method relies on determining the fields in the region between the two metasurfaces, so that the power density is matched at both metasurfaces simultaneously. In Sec.~\ref{Sec:MOM}, we reformulate the theoretical framework by expanding the fields at the inner boundaries of the metasurface pair into two summations of spatially-shifted basis functions that propagate in opposite directions. With this choice of basis functions expansion in the spatial domain, we avoid the discretization of fields in terms of modes (wavevectors) and angles of propagation, as done in \cite{Ayman:AWPL2018}. In addition, the use of two counter-propagating field distributions allows for easily introducing multiple reflections in the design. The proposed design method can be adjusted to handle two geometry variations, depending on whether the source is placed outside or within the metasurface pair. To validate our formulation and suggest possible applications, we present several examples of wave transformations that break the local power conservation condition in Sec.~\ref{Sec:DesignExamples}. For the first geometry variation with the source outside the metasurface pair, we design a cylindrical-wave to plane-wave transformation and a Gaussian beam expander. Through these examples, we examine the usefulness of multiple reflections in reducing the separation distance between the two metasurfaces. For the second scenario with the source located within the metasurface pair, we show that both sides can be treated as independent output apertures and the incident power from the source can be arbitrarily splitted between them. Based on this configuration, we present a beamforming example featuring a single-sided or a double-sided Taylor distribution output. In Sec.~\ref{Sec:CavityEffects}, we comment on the influence of cavity effects to the sensitivity of the proposed design and we investigate possible ways to render it less susceptible to geometrical and, indirectly, frequency variations. Lastly, we conclude our work in Sec.~\ref{Sec:Conclusion}.

\section{Method of moments approximation}
\label{Sec:MOM}
\begin{figure}
\centering
{\includegraphics[width=\columnwidth, trim=0 0 0 0 0, clip=true]{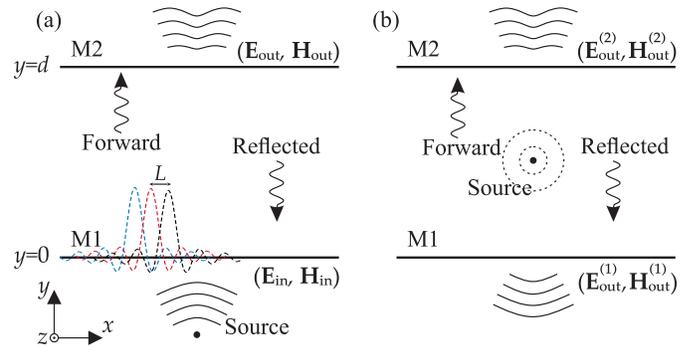}}
\caption{Two lossless passive omega-bianisotropic Huygens' metasurfaces that are used for arbitrary field transformations. (a) The source is placed before the first metasurface. (b) The source is placed between the two metasurfaces.}
\label{fig:Fig1}
\end{figure}

In this section, the theoretical formulation for designing a pair of passive and lossless omega-bianisotropic HMSs is developed. Two configurations are considered depending on the location of the source, as illustrated in Fig.~\ref{fig:Fig1}. In the first one, the electromagnetic source is placed before the first metasurface (M1) and the input field distribution $\{ \mathbf{E}_\mathrm{in}, \mathbf{H}_\mathrm{in} \}$ is transformed to an arbitrary output field distribution $\{ \mathbf{E}_\mathrm{out}, \mathbf{H}_\mathrm{out} \}$ at the output of the second metasurface (M2). On the contrary, in the second configuration the source is placed between the two metasurfaces and two arbitrary output field distributions, represented by $\{ \mathbf{E}^{(1)}_\mathrm{out}, \mathbf{H}^{(1)}_\mathrm{out} \}$ and $\{ \mathbf{E}^{(2)}_\mathrm{out}, \mathbf{H}^{(2)}_\mathrm{out} \}$, can be supported at the two sides of the metasurface pair. For both geometries, it is assumed that the metasurfaces extend infinitely along the $z$-direction for simplicity. The distance between them is denoted by $d$, while $L_\mathrm{tot}$ stands for the total width of each metasurface. Furthermore, the electric field is assumed to be transverse electric (TE) polarized with the only non-zero component being along $\hat{\mathbf{z}}$.

\subsection{Calculation of real power at the boundaries of the two metasurfaces}

As observed in Fig.\ref{fig:Fig1}, the fields at the outer boundaries of the metasurface pair (i.e., $y=0^-$ for M1 and $y=d^+$ for M2) are known for both configurations, as long as the input fields (or the type of the source) and the desired output fields are specified. Therefore, the real power density flowing normal to the metasurfaces at the planes $y=0^-$ and $y=d^+$ can be calculated through the Poynting vector. Then, the aim is to properly define the electromagnetic fields between the two metasurfaces in order to satisfy local power conservation at both boundaries simultaneously, while obeying Maxwell's equations. To this purpose, two field distributions are introduced propagating to the forward ($+\hat{\mathbf{y}}$) and to the backward ($-\hat{\mathbf{y}}$) directions. When the source is placed before M1, the summation of the two counter-propagating field distributions equals the total fields between the two metasurfaces, whereas the incident fields from the source should also be added in case that it is located within the metasurface pair.

The electric field of the forward-propagating wave $E^\mathrm{forw}_{z}(x,y)$ is approximated at M1 ($y=0^+$) as a summation of $2N+1$ terms
\begin{align}
\label{eq:E_expansion}
E^\mathrm{forw}_z (x,y=0^+)= \sum_{n=-N}^{n=N}  A_{f,n} g_n(x),
\end{align}
where $A_{f,n}$ are the unknown complex weights, still to be determined. The basis functions $g_n(x)$ in Eq.~\eqref{eq:E_expansion} are defined as
\begin{align}
\label{eq:basis_func}
g_n(x)=\frac{\mathrm{sin}\left(k(x-nL)\right)}{k \left(x-nL\right)},
\end{align}
where $k=2\pi/\lambda$ is the free-space wave-vector ($\lambda$ standing for the wavelength of operation) and $L=L_\mathrm{tot}/(2N+1)$ represents the spatial shift between two adjacent functions $g_n(x)$, as illustrated in Fig.~\ref{fig:Fig1}(a). By taking the Fourier transform of Eq.~\eqref{eq:E_expansion} with respect to the $x$-coordinate, we can write the spatial spectrum of the forward-propagating electric field after the first metasurface
\begin{align}
\label{eq:FT_expansion}
\hspace{-0.2em}E^\mathrm{forw}_{z}(k_x, y=0^+)\hspace{-0.2em}=\hspace{-0.2em}
\begin{dcases} 
\frac{\pi}{k} \sum_{n=-N}^{n=N} A_{f,n} e^{+j k_x nL} &\hspace{-0.6em}, |k_x| \leq k \\
0 & \hspace{-0.6em},|k_x| > k.
\end{dcases}
\end{align} 
From the selection of the basis functions in Eq.~\eqref{eq:E_expansion}, it is evident that only the propagating part ($|k_x| \leq k$) of the spectrum is considered. However, in the case of relatively smooth input and output power distributions the absence of evanescent waves does not restrict the ability to achieve power matching at the two metasurfaces. Each of the spectral components of the wave propagates along $+\hat{\mathbf{y}}$ with a wavenumber $k_y=\sqrt{k^2-k_x^2}$. Subsequently, the  forward-propagating electric field at any given $y$-plane is given as the following inverse Fourier transform
\begin{align}
\label{eq:Efield_forw}
E^\mathrm{forw}_{z}&(x,y)=\frac{1}{2 \pi} \int_{-k}^k E^\mathrm{forw}_{z}(k_x,y=0^+) e^{-jk_yy} e^{-jk_xx}  dk_x \nonumber \\
&= \sum_{n=-N}^{n=N} \frac{A_{f,n}}{2 k}  \int_{-k}^k e^{-j \sqrt{k^2-k_x^2}y} e^{-jk_x(x-nL)} dk_x.
\end{align}
From Maxwell's equations, the magnetic field component tangential to the metasurface can be calculated as
\begin{align}
\label{eq:Hfield_forw}
& H^\mathrm{forw}_x(x,y)=\frac{j}{\omega \mu} \frac{\partial E^\mathrm{forw}_z}{\partial y} \nonumber \\
&= \sum_{n=-N}^{n=N} \frac{A_{f,n}}{2  k^2 \eta}  \int_{-k}^k \sqrt{k^2-k_x^2} e^{-j \sqrt{k^2-k_x^2}y} e^{-jk_x(x-nL)} dk_x,
\end{align}
where $\eta=\sqrt{\mu/\varepsilon}$ is the characteristic impedance of the medium.

Likewise, the reflected wave is expanded at the second metasurface ($y=d^-$) as
\begin{align}
\label{eq:E_expansion_ref}
E^\mathrm{ref}_{z}(x,y=d^-)= \sum_{n=-N}^{n=N}  A_{r,n} g_n(x),
\end{align}
where the complex weights $A_{r,n}$ form another set of unknowns. By following a similar approach for the reflected fields (defining in this case the zero-phase plane at $y=d$ and propagating each Fourier component at the $-\hat{\mathbf{y}}$ direction), we arrive at the following expression for the reflected electric field at any $y$-plane
\begin{align}
\label{eq:Efield_ref}
E^\mathrm{ref}_{z}&(x,y)= \sum_{n=-N}^{n=N}\frac{A_{r,n}}{2 k}  \int_{-k}^k e^{j \sqrt{k^2-k_x^2}(y-d)} e^{-jk_x(x-nL)} dk_x.
\end{align}
Then, the tangential component of the reflected magnetic field is obtained directly from Maxwell's equations as
\begin{align}
\label{eq:Hfield_ref}
& H^\mathrm{ref}_x(x,y)= \nonumber \\
& -\sum_{n=-N}^{n=N} \frac{A_{r,n}}{2  k^2 \eta}  \int_{-k}^k \sqrt{k^2-k_x^2} e^{j \sqrt{k^2-k_x^2}(y-d)} e^{-jk_x(x-nL)} dk_x.
\end{align}

As stated above, in the case that the source is placed before M1, as shown in Fig.~\ref{fig:Fig1}(a), the total electric and magnetic fields between the two metasurfaces are found as a superposition of the forward wave in Eqs.~\eqref{eq:Efield_forw}-\eqref{eq:Hfield_forw} and the reflected wave in Eqs.~\eqref{eq:Efield_ref}-\eqref{eq:Hfield_ref}. By substituting $y=0$ and $y=d$ in the above expressions, the total tangential electric and magnetic fields are calculated at the inner boundaries of the two metasurfaces
\begin{subequations}\label{eqs:field_expressions}
\begin{align}
E_{z}(x,y=0^+)&= \sum_{n=-N}^N A_{f,n} I_n + \sum_{n=-N}^N A_{r,n} I_n', \\
H_{x}(x,y=0^+)&= \sum_{n=-N}^N A_{f,n} J_n - \sum_{n=-N}^N A_{r,n} J_n', \\
E_{z}(x,y=d^-)&= \sum_{n=-N}^N A_{f,n} I_n' + \sum_{n=-N}^N A_{r,n} I_n, \\
H_{x}(x,y=d^-)&= \sum_{n=-N}^N A_{f,n} J_n' - \sum_{n=-N}^N A_{r,n} J_n,
\end{align}
\end{subequations}
where $I_n, I_n',J_n, J_n'$ are coefficients that depend on the $x$-coordinate according to the expressions
\begin{subequations}\label{eq:Coefficients}
\begin{align}
&I_n=\frac{1}{2 k} \int_{-k}^k  e^{-jk_x(x-nL)} dk_x, \\
&I_n'=\frac{1}{2 k} \int_{-k}^k e^{-j \sqrt{k^2-k_x^2}d} e^{-jk_x(x-nL)} dk_x, \\
&J_n=\frac{1}{2 k^2 \eta} \int_{-k}^k \sqrt{k^2-k_x^2}  e^{-jk_x(x-nL)} dk_x, \\
&J_n'=\frac{1}{2 k^2 \eta} \int_{-k}^k \sqrt{k^2-k_x^2} e^{-j \sqrt{k^2-k_x^2}d} e^{-jk_x(x-nL)} dk_x. 
\end{align}
\end{subequations}
The real power density is then determined as
 \begin{subequations}\label{eq:Powers}
\begin{align}
&P_y(x,y=0^+)= \frac{1}{2} \mathrm{Re} \Bigg\{ \sum_{n=-N}^{n=N} \sum_{m=-N}^{m=N} \Big(I_n J_m^* A_{f,n} A_{f,m}^*+ \nonumber \\
&I_n' J_m^* A_{r,n} A_{f,m}^* - I_n J_m^{'*} A_{f,n} A_{r,m}^* -I_n' J_m^{'*} A_{r,n} A_{r,m}^* \Big)\Bigg\}, \\
&P_y(x,y=d^-)= \frac{1}{2} \mathrm{Re} \Bigg\{ \sum_{n=-N}^{n=N} \sum_{m=-N}^{m=N} \Big(I_n' J_m^{'*} A_{f,n} A_{f,m}^*+ \nonumber \\
&I_n J_m^{'*} A_{r,n} A_{f,m}^* - I_n' J_m^{*} A_{f,n} A_{r,m}^* -I_n J_m^* A_{r,n} A_{r,m}^* \Big)\Bigg\}.
\end{align}
\end{subequations}
For a given pair of metasurfaces (fixed separation $d$, width $L_\mathrm{tot}$ and discretization $N$), the coefficients in Eqs.~\eqref{eq:Coefficients} can be numerically determined and the real power density, calculated in Eqs.~\eqref{eq:Powers}, solely depends on the two sets of unknown weights $A_{f,n}$ and $A_{r,n}$.

On the other hand, when the source is placed between the two metasurfaces, the total field expressions in Eqs.~\eqref{eqs:field_expressions} should be supplemented by the incident fields produced directly from the source. The expressions for the real power density at the two boundaries are, then, modified to
\begin{subequations}\label{eq:Powers_modified}
\begin{align}
P_{y}(x,y=&0^+)= \frac{1}{2} \mathrm{Re} \Bigg\{ \sum_{n=-N}^{n=N}   \Big( I_n A_{f,n}  + I_n' A_{r,n} + E_{\mathrm{inc},z}^{(1)} \Big) \nonumber \\
&   \sum_{m=-N}^{m=N} \Big( J_m A_{f,m}  - J_m' A_{r,m} + H_{\mathrm{inc},x}^{(1)} \Big)^* \Bigg\}, \\
P_{y}(x,y=&d^-)= \frac{1}{2} \mathrm{Re} \Bigg\{ \sum_{n=-N}^{n=N}   \Big( I_n' A_{f,n}  + I_n A_{r,n} + E_{\mathrm{inc},z}^{(2)} \Big) \nonumber \\
&   \sum_{m=-N}^{m=N} \Big( J_m' A_{f,m}  - J_m A_{r,m} + H_{\mathrm{inc},x}^{(2)} \Big)^* \Bigg\},
\end{align}
\end{subequations}
where $\{E_{\mathrm{inc},z}^{(1)}, H_{\mathrm{inc},x}^{(1)} \}$ and $\{E_{\mathrm{inc},z}^{(2)}, H_{\mathrm{inc},x}^{(2)} \}$ are the profiles of the incident tangential electric and mangetic fields at M1 and M2, respectively, as calculated from the source in the absence of the two metasurfaces.

Regarding the configuration in Fig.~\ref{fig:Fig1}(a) with the source located before M1, the use of two counter-propagating waves allows for the handling of multiple reflections between the two metasurfaces. Certainly, in the case that the weights of the reflected wave $A_{r,n}$ are set to zero, multiple reflections are not considered and the field transformation is designed to be performed by two purely transmissive omega-bianisotropic Huygens' metasurfaces, as previously shown in \cite{Raeker:PRL2019,Ayman:AWPL2018}. However, the hypothesis here is that the use of multiple reflections between the two metasurfaces allows to reduce the separation of the two metasurfaces for a desired field transformation. This is theoretically justified, since the propagation between the two metasurfaces is the mechanism to redistribute the power; thus, multiple reflections effectively increase the propagation length, while maintaining the compactness of the design. Needless to say, the use of a backward-propagating wave is also advantageous in the second configuration of Fig.~\ref{fig:Fig1}(b), as multiple reflections allow for more accurate redistribution of the total power and improved local power matching at the two sides of each metasurface.

Another important aspect is specifying the level of the total output power so that maximum power efficiency is ensured for the designed metasurface pair. Since the metasurfaces are considered lossless, the only loss mechanism is the power escaping from the open sides of the structure. To minimize this leakage, the total output power $P_\mathrm{out}^\mathrm{tot}$ is made equal with the total incident power $P_\mathrm{inc}^\mathrm{tot}$ from the source, which is
\begin{align} \label{eq:Pav1}
P_\mathrm{inc}^\mathrm{tot} =  \frac{1}{2} \int_{M1} \mathrm{Re} \{  E_{\mathrm{in},z}  H_{\mathrm{in},x}^* \} dx,
\end{align}
or 
\begin{align} \label{eq:Pav2}
P_\mathrm{inc}^\mathrm{tot} = & - \frac{1}{2}  \int_{M1}   \mathrm{Re} \{ E_{\mathrm{inc},z}^{(1)}  H_{\mathrm{inc},x}^{(1)*} \}dx \nonumber \\
& + \frac{1}{2} \int_{M2} \mathrm{Re} \{ E_{\mathrm{inc},z}^{(2)}  H_{\mathrm{inc},x}^{(2)*} \} dx,
\end{align}
when the source is placed outside or inside the metasurface pair, respectively. The incident power represents the part of the input power that can be handled by the metasurface pair, as it is not directly leaked outside of the configuration.

\subsection{{Point matching of the real power}}
\label{Sec:Point_matching}
To achieve local power conservation at each metasurface, the input/output power densities at $y=0^-$ and $y=d^+$ are sampled at the locations $x=nL, n=-N,...,N$ and they are equated with the power densities at the inner boundaries of the two metasurfaces at $y=0^+$ and $y=d^-$, respectively, as given by Eqs.~\eqref{eq:Powers} or Eqs.~\eqref{eq:Powers_modified}. Conceptually, this process is equivalent to point matching of the power density at two sets of equidistant points along the two metasurfaces. However, unlike the classic Method of Moments formulation, the electromagnetic quantity involved in the point matching is the power density instead of the electric field values. As a result of the point-matching process, a system of $2(2N+1)$ equations is formed as
\begin{align}
\mathbf{G}=
\begin{pmatrix*}[l]
P_{y}(y=0^-)-P_{y}(y=0^+)\Big|_{x=-NL} \\
P_{y}(y=0^-)-P_{y}(y=0^+)\Big|_{x=-(N-1)L} \\
... \\
P_{y}(y=0^-)-P_{y}(y=0^+)\Big|_{x=NL} \\
P_{y}(y=d^-)-P_{y}(y=d^+)\Big|_{x=-NL} \\
P_{y}(y=d^-)-P_{y}(y=d^+)\Big|_{x=-(N-1)L} \\
... \\
P_{y}(y=d^-)-P_{y}(y=d^+)\Big|_{x=NL} \\
\end{pmatrix*} = 0,
\end{align}
which should be solved for the unknown complex weights $A_{f,n}$ and $A_{r,n}$, that determine the power densities at the inner boundaries of the metasurface pair.

Due to the quadratic nature of the expressions in Eqs.~\eqref{eq:Powers}-\eqref{eq:Powers_modified}, the system is nonlinear and a gradient descent method is utilized to minimize the total power mismatch at the two metasurfaces. More specifically, the real and the imaginary parts of each weight define the vector of the unknowns, denoted by $\mathbf{x}$, that is optimized in order to minimize the objective function
\begin{align}
F(\mathbf{x})=\frac{1}{2} \mathbf{G}^T \mathbf{G}.
\end{align}
The Jacobean matrix $\mathbf{J}_G$, involving the derivatives of each row of the $\mathbf{G}$ vector with respect to each unknown is analytically calculated (as a function of the coefficients $I_n,I_n',J_n,J_n'$). Then, at every iteration the vector of the unknowns is updated based on the expression
\begin{align}
\mathbf{x}^{(n+1)}=\mathbf{x}^{(n)}-\gamma \mathbf{J}_G (\mathbf{x}^{(n)})^T\mathbf{G}(\mathbf{x}^{(n)}),
\end{align}
where $\mathbf{x}^{(n)}$ is the vector of the unknowns at the $n$-th iteration and $\gamma$ represents the learning rate of the optimization algorithm, which is carefully selected for each design problem.

\subsection{Metasurface macroscopic design}

Once the optimization algorithm has converged to a solution that minimizes the local power mismatch at the two metasurfaces, the tangential fields at the two sides of each metasurface are used for their design. Due to local power conservation, the field transformations at the two boundaries are guaranteed to be possible with reflectionless and lossless omega-bianisotropic Huygens' metasurfaces \cite{Epstein:TAP2016}. Moreover, the parameters of each metasurface, namely the surface electric impedance $Z_{se}$, the surface magnetic susceptibility $Y_{sm}$ and the magnetoelectric coupling coefficient $K_{em}$, can be uniquely determined at every sampling point through \cite[Eq.~(5)]{Epstein:TAP2016}. It is emphasized that the above-cited expressions for designing a passive and lossless metasurface require the local power conservation condition to be satisfied perfectly. However, they are still applicable for the fields obtained from the optimization method discussed in Sec.~\ref{Sec:Point_matching}, as long as the power mismatch at the two boundaries is relatively low. It is also worth mentioning that a constant phase can be added to both the tangential electric and magnetic fields at the output of each metasurface. While adding a constant phase does not affect the power density matching, it can be advantageous to avoid extreme values for the metasurface parameters and facilitate the convergence of the simulations.

Full-wave simulations are performed in ANSYS HFSS by realizing the metasurfaces with a three-layer impedance sheet structure, consisting of three lossless impedance layers, whose values are given by \cite[Eq.~(8)]{Epstein:TAP2016}. The layers in our simulations are separated by extremely thin ($\lambda/800$) air  regions, as this was beneficial to handle some convergence issues of the full-wave simulations of impedance sheets in ANSYS HFSS. In practical designs, these abstract impedance layers are to be replaced by copper traces etched on standard substrates and bonded together. The coupling between the layers, as well as the copper and dielectric losses may be significant and should be taken into account before the fabrication of the metasurfaces \cite{Chen:PRB2018}. However, our purpose here is only to validate the proposed theoretical framework through full-wave simulations; hence, the three cascaded impedance layers are a sufficient model for the unit cells of the metasurfaces. The simulation domain is confined within two parallel perfect electric conducting (PEC) plates in the $z$-axis to guarantee uniformity along this direction and TE-polarized fields, while the lateral sides are terminated with perfectly-matched-layer (PML) boundaries.

\section{Design examples}
\label{Sec:DesignExamples}
\subsection{Uniform aperture illumination by a single line-source excitation} \label{Sec:UniformOutput}

In the first example, the configuration described in Fig.~\ref{fig:Fig1}(a) is used to transform an incident cylindrical wave to a uniform field distribution with constant phase along the output aperture. For a given aperture length, the uniform distribution exhibits the maximum directivity and the minimum half-power beamwidth (HPBW). An infinite (along the $z$-axis) current line-source operating at the frequency $f=10 \ \mathrm{GHz}$ is placed at distance $s=\lambda/3$ before M1, where $\lambda$ is the respective free-space wavelength. The current source radiates a cylindrical wave that partially impinges on the first metasurface, having the following profile for the tangential electric and magnetic fields,
\begin{subequations}\label{eq:Incident_Plane}
\begin{align}
& E_{\mathrm{in},z}(x)=-\frac{k \eta I}{4} H_0^{(2)}(k\sqrt{x^2+s^2}),\\
& H_{\mathrm{in},x}(x)= \frac{j k I}{4} \frac{s}{\sqrt{x^2+s^2}} H_1^{(2)}(k\sqrt{x^2+s^2}), 
\end{align}
\end{subequations}
where $H_{n}^{(2)}$ is the $n$-th order Hankel function of the second kind and $I=1 \ \mathrm{A}$ is an arbitrarily chosen current amplitude. The purpose of the designed metasurface pair is to transform the fields into a truncated plane wave, characterized by a constant amplitude along the output aperture. Both metasurfaces have a width of $L_\mathrm{tot}=6 \lambda$ and are discretized with 101 unit cells  ($N=50$).

\begin{figure}
\centering
{\includegraphics[width=0.85\columnwidth, trim=0 0 0 0 0, clip=true]{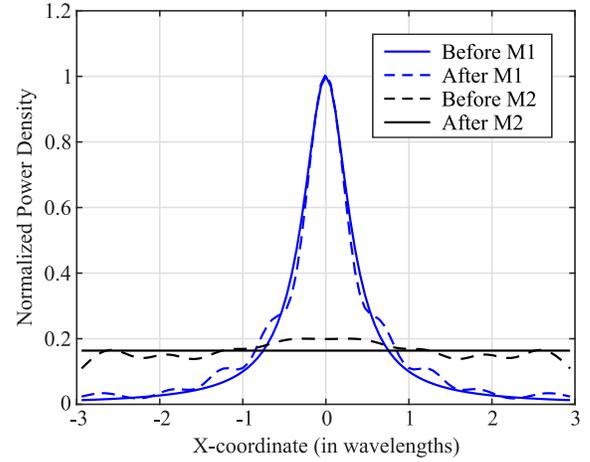}}
\caption{Local power matching at the two metasurfaces as given by the optimization method for $d=1.5 \lambda$. The power densities at the inner sides of the configuration (dashed lines) should locally match the defined input and output power distributions (solid lines) at both metasurfaces.}
\label{fig:Fig2}
\end{figure}

\begin{figure}
\centering
{\includegraphics[width=0.8\columnwidth, trim=0 0 0 0 0, clip=true]{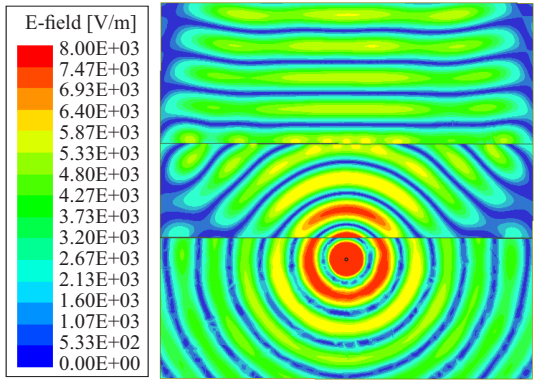}}
\caption{Real part (absolute values) of the electric field $E_z$ for a separation distance $d=1.5\lambda$ and totally reflectionless metasurfaces. The cylindrical wave (bottom region) transforms to a truncated plane wave propagating at the output (upper region) after passing through the metasurface pair.}
\label{fig:Fig3}
\end{figure}

To verify the applicability of our method, we first design a pair of reflectionless (i.e. $A_{r,n}=0$, for all $n$) metasurfaces with a separation distance of $d=1.5 \lambda$. Even without multiple reflections, this separation distance is sufficient to achieve acceptable power matching at both boundaries simultaneously. The fields between the two metasurfaces are calculated using the optimization method described in Sec.~\ref{Sec:Point_matching} and the normal real power density at the two sides of each metasurface is depicted in Fig.~\ref{fig:Fig2}. Since the design process of the two omega-bianisotropic Huygens' metasurfaces assumes perfect local power conservation, it is expected that any deviations at either of the two metasurfaces will induce reflections before M1 and perturb the transmitted fields at the output of M2. The real part (in absolute values) of the electric field is depicted in Fig.~\ref{fig:Fig3}, where it is clear that the wavefronts transform from cylindrical at the input to planar at the output. Moreover, the amplitude of the electric field at the output of M2 is kept nearly constant with the fluctuations attributed to the power mismatch at the two metasurfaces.

\begin{figure}
\centering
{\includegraphics[width=0.9\columnwidth, trim=0 0 0 0 0, clip=true]{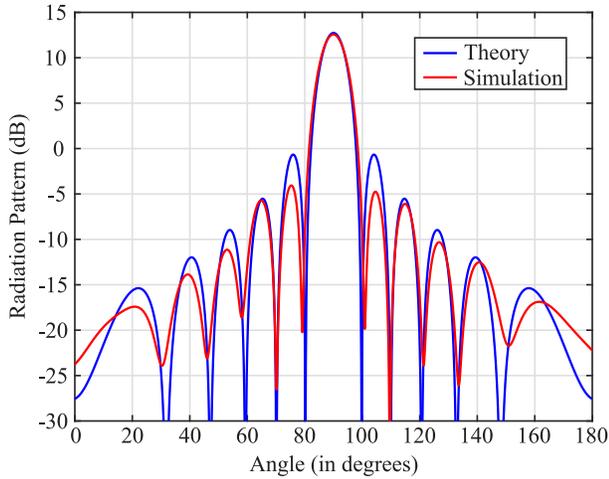}}
\caption{Radiation pattern in the forward direction. The simulation results (red line) match the theoretical radiation pattern of a uniform aperture (blue line).}
\label{fig:Fig4}
\end{figure}

It should be noted again that the sides between the two metasurfaces are open boundaries that do not confine the electromagnetic power within the width of the metasurface pair. However, by defining the input and the output power distributions to have the same total power, the real power leaking to the two sides is minimized and it is solely attributed to the small power mismatch at the two metasurfaces. In the case considered $95.2 \% $ of the incident power, as defined in Eq.~\eqref{eq:Pav1}, is transmitted to the output of the second metasurface, while only $0.6 \%$ is reflected and the rest escapes from the two sides. The radiation pattern is depicted in Fig.~\ref{fig:Fig4} together with the theoretical one for a perfectly uniform output distribution. For the calculation of the radiation pattern, only the boundaries after M2 ($y>d$) are considered for the near-field to far-field transformation. It is observed that the directivity and the beamwidth compare well with the predicted values. Specifically, the obtained directivity from the simulation is $12.57 \ \mathrm{dB}$ that is very close to the theoretical value of $12.73 \ \mathrm{dB}$ for an aperture width of $6 \lambda$. Regarding the main lobe beamwidth, a HPBW of $9.2\degree$ is calculated that is relatively close to the predicted value of $8.3\degree$ for a uniformly excited aperture.

\begin{figure}
\centering
{\includegraphics[width=0.8\columnwidth, trim=0 0 0 0 0, clip=true]{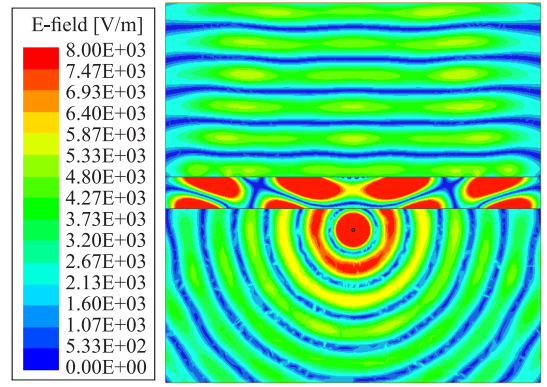}}
\caption{Real part (absolute values) of the electric field $E_z$ for a separation distance of $d=0.5\lambda$ and utilizing multiple reflections between the two metasurfaces.}
\label{fig:Fig5}
\end{figure}

\begin{figure}
\centering
{\includegraphics[width=0.9\columnwidth, trim=0 0 0 0 0, clip=true]{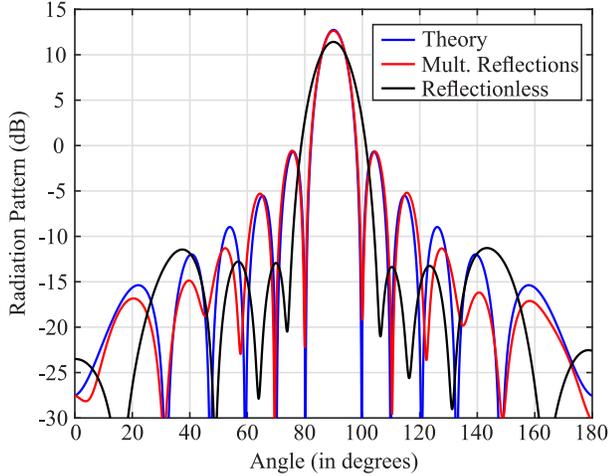}}
\caption{Radiation pattern for a separation distance of $d=0.5\lambda$ with multiple reflections (red line) and with a reflectionless design (black line). The use of multiple reflections enhances the quality of the transformation and the obtained radiation pattern matches better with the theoretical (blue line).}
\label{fig:Fig6}
\end{figure}

Further reducing the separation between the two metasurfaces leads to higher power mismatch and gradually degrades the total performance of the field transformation. These effects can be mitigated by allowing multiple reflections between the two metasurfaces. Although the total configuration remains reflectionless, multiple reflections effectively increase the propagation length and lead to better power matching, especially for smaller separation lengths. The same field transformation is examined with a distance $d=0.5\lambda$ between the two metasurfaces and the use of multiple reflections. The designed pair of metasurfaces is simulated and the electric field is plotted in Fig.~\ref{fig:Fig5}. Standing-wave patterns with higher field values can be observed in the region between the two metasurfaces, since these are partially reflective and form a cavity (open to the sides) between them. The electric field at the output aperture still exhibits nearly uniform amplitude and phase despite the reduced size of the metasurface pair. To better evaluate the usefulness of multiple reflections at reducing the size of our design, the reflectionless scenario is also simulated and the two cases are compared in terms of the far-field characteristics. The radiation pattern is plotted in Fig.~\ref{fig:Fig6}, where it is evident that the use of multiple reflections is essential to get nearly uniform aperture illumination at the output. When multiple reflections are present, the obtained directivity is $12.64 \ \mathrm{dB}$, the HPBW is $8.8\degree$ and the sidelobe level is $-13.2 \ \mathrm{dB}$, which are all very close to the theoretical values for a uniformly illuminated aperture. On the contrary, in the totally reflectionless case the directivity is reduced to $11.42 \ \mathrm{dB}$, while the HPBW is increased to $12.3\degree$. Defining the aperture efficiency as the ratio of the obtained maximum directivity divided by the theoretical value of a uniform aperture, this results in only $74 \%$ in the totally reflectionless case compared to $98 \%$ aperture efficiency with the use of multiple reflections in the design.

\subsection{Beam expansion for an incident Gaussian beam} \label{Sec:BeamExpander}
Beam expanders are commonly used in optics to alter the width of collimated light beams \cite{Smith:Book}. In their traditional form, they consist of two dielectric lenses of appropriate focal lengths to provide the required scaling to the beam radius. Recently, a pair of omega-bianisotropic Huygens' metasurfaces has been employed at microwaves to design a beam expander \cite{Chen:Metamaterials2018}. In \cite{Chen:Metamaterials2018} both metasurfaces are purely transmissive and act on the incident fields by adding an appropriate phase profile, so that the first serves as a diverging lens and the second as a converging lens with the designed focal lengths. However, this approach has limitations regarding the separation between the two metasurfaces for a desired scaling factor of the beam radius; as the distance between the two metasurfaces becomes smaller, the performance of the transformation is severely degraded. To overcome these limitations, multiple reflections can be utilized, as suggested in our previous example.

In this example, the two metasurfaces are separated by a distance of $d=0.5 \lambda$, where $\lambda \approx 30 \ \mathrm{mm}$ is the wavelength at the frequency of $10 \ \mathrm{GHz}$. Both metasurfaces are $L_\mathrm{tot}=9\lambda$ wide and are discretized with $101$ unit cells (resulting approximately at a $\lambda/11$ unit cell size). The input wave is assumed to be a Gaussian beam with its focus at M1; thus, the electric field at M1 is the following
\begin{align} \label{eq:Gaussian_Ein}
E_{\mathrm{in},z}(x)=A_\mathrm{in} \mathrm{exp} \{-x^2/w_\mathrm{in}^2 \},
\end{align}  
where $w_\mathrm{in}=1.5\lambda$ is the beam waist and $A_\mathrm{in}=10^3 \ \mathrm{V}/\mathrm{m}$ is a arbitrarily chosen peak amplitude. By decomposing the electric field in Eq.~\eqref{eq:Gaussian_Ein} to its plane-wave components, computing the magnetic field for each one of them and summing the individual contributions, we can calculate the total magnetic field at M1 as
\begin{align} \label{eq:Gaussian_Hin}
H_{\mathrm{in},x}(x) = \frac{1}{2 \pi} \int_{-k}^{k} \frac{k_y}{\eta k} \left[ \int_{-L_{\mathrm{tot}}/2}^{L_{\mathrm{tot}}/2} E_{\mathrm{in},z}(x) e^{j k_x x} dx \right] dk_x.
\end{align}
As in the case of the unknown forward and reflected waves, it is assumed that the field spectrum is sufficiently decayed for $|k_x|>k$ and the integration in Eq.~\eqref{eq:Gaussian_Hin} is constrained only to the propagating plane-wave components. The aim is to get an output Gaussian beam with a waist of $w_\mathrm{out}=3\lambda$ at M2, while conserving the total propagating power. The output tangential fields at M2 are specified similar to Eqs.~\eqref{eq:Gaussian_Ein}-\eqref{eq:Gaussian_Hin} with the output beam waist $w_\mathrm{out}$ and an appropriately chosen amplitude $A_\mathrm{out}$ to ensure that the total real power for the two distributions is conserved. 

First, we design the beam expander based on two independently defined metasurfaces that add a phase profile on the transmitted fields, the first acting as a diverging lens with focal length $f_1=-d$ and the second as a converging lens with focal length $f_2=2d$. Secondly, we compare this approach with the proposed optimization method including multiple reflections in the formulation. The use of multiple reflections allows for better power matching at the two metasurfaces, as expected from the higher number of degrees of freedom involved in the optimization process. Consequently, the field transformation is expected to be more accurate for the same separation distance, compared to the case of two totally reflectionless metasurfaces that act solely on the phase similar to traditional lenses.
\begin{figure}
\centering
{\includegraphics[width=0.8\columnwidth, trim=0 0 0 0 0, clip=true]{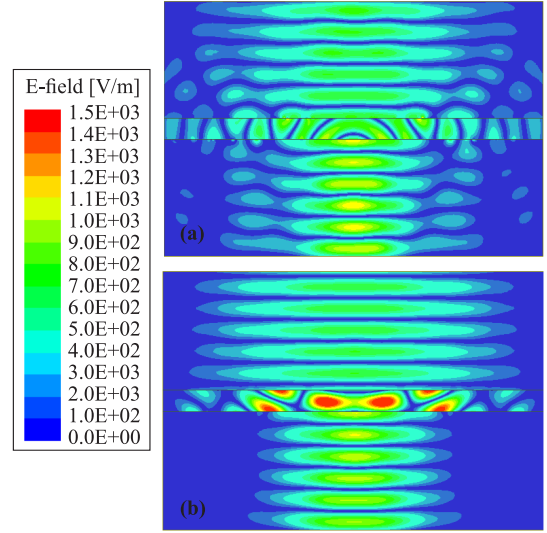}}
\caption{Real part (absolute values) of the electric field $E_z$ for the Gaussian beam expander. (a) The two HMSs are designed independently as lenses that only add a phase profile in the transmitted fields. (b) The two HMSs are optimized based on the proposed method and multiple reflections between them are considered. The distance between the two HMSs is $d=0.5 \lambda$ for both designs.}
\label{fig:Fig7}
\end{figure}
Both designs are simulated and the real part of the electric field is depicted in Fig.~\ref{fig:Fig7}. It is clearly observed that using purely transmissive phase-changing Huygens' metasurfaces results in considerable deviations from the desired output fields and reflections that perturb the fields in the input region before M1. The same issues are also present for this distance between the two metasurfaces, if the proposed optimization method is used but without introducing multiple reflections. On the contrary, the above-mentioned problems are substantially mitigated when multiple reflections are considered, as it is evident from Fig.~\ref{fig:Fig7}(b).

\begin{figure}
\centering
{\includegraphics[width=0.9\columnwidth, trim=0 0 0 0 0, clip=true]{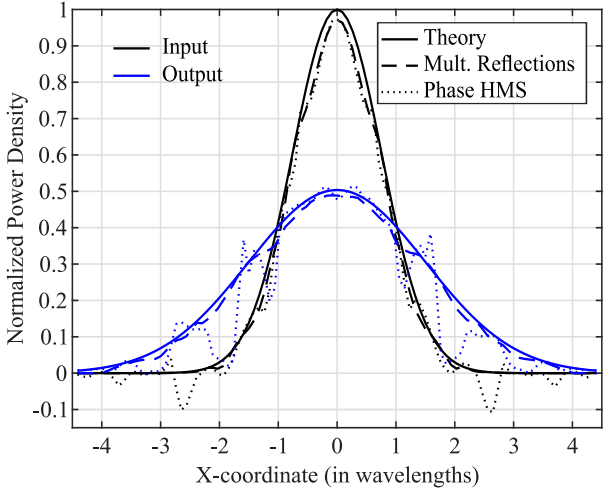}}
\caption{Input (black curves) and output (blue curves) normalized power density profiles for the Gaussian beam expander. The design including multiple reflections (dashed lines) matches better with the expected theoretical values (solid lines) compared to the simulation of two phase-changing metasurfaces acting as lenses (dotted lines).}
\label{fig:Fig8}
\end{figure}

To better evaluate the accuracy of the field transformation, the real power density is plotted at two cuts of the configuration; the first at a distance $\lambda/10$ before the first metasurface and the second $\lambda/10$ after the second metasurface. This small offset is introduced so that any rapid field oscillations close to the metasurfaces owing to the discretization and the discontinuity of the fields at the boundary are sufficiently decayed. The normalized power densities are plotted together with the theoretical input and output power distributions in Fig.~\ref{fig:Fig8}. Both the input and the output power density profiles are close to the theoretical ones for the design that includes multiple reflections between the two metasurfaces. This suggests that the desired field transformation is successfully performed for this case, while noticeable deviations exist in the design involving phase-changing reflectionless metasurfaces. Lastly, it should be noted that the power efficiency with the proposed design method is $94 \%$ with only $6 \%$ of the incident power being reflected or escaping through the two sides.

\subsection{Single-sided beamforming with a single line-source in the middle of the metasurface pair} \label{sec:Single-sided Taylor}
In various antenna applications, radiation patterns with specific characteristics (directivity, side lobe level, main lobe beamwidth, etc.) are required. To this purpose, metasurfaces have been used in different configurations to obtain the aperture fields at the output that would produce the desirable radiation pattern in the far-field region \cite{Epstein:TAP2017Antennas,Raeker:PRL2019}. In the following, we demonstrate how the proposed design of a pair of omega-bianisotropic Huygens' metasurfaces can be utilized for the realization of a low-profile Taylor (one parameter) antenna that is excited by a single current line-source placed within the metasurface pair, as shown in Fig.~\ref{fig:Fig1}(b). The metasurfaces are $L_\mathrm{tot}=6\lambda$ wide ($\lambda$ being the free-space wavelength), the operating frequency is $f=10 \ \mathrm{GHz}$ and the separation distance is set to $d=0.75\lambda$. The current line-source is placed exactly in the middle of M1 and M2 ($y=d/2$) and radiates a cylindrical wave that has the following tangential components at the two metasurfaces
\begin{subequations}\label{eq:Incident_Taylor}
\begin{align}
& E^{(p)}_{\mathrm{inc},z}(x)=-\frac{k \eta I}{4} H_0^{(2)}(k\sqrt{x^2+(d/2)^2}),\\
& H^{(p)}_{\mathrm{inc},x}(x)=\mp \frac{j k I d}{4 \sqrt{4x^2+d^2}} H_1^{(2)}(k\sqrt{x^2+(d/2)^2}), 
\end{align}
\end{subequations}
where $I=1 \ \mathrm{A}$ is the current amplitude and $p=\{1,2\}$ refers to the fields at the corresponding metasurface and the minus (plus) sign is taken in Eq.~(\ref{eq:Incident_Taylor}b) for $p=1$ ($p=2$).

Since a single output is desired, the fields $\{ \mathbf{E}^{(1)}_\mathrm{out}, \mathbf{H}^{(1)}_\mathrm{out} \}$ below M1 are set to zero. On the contrary, the electric field at the upper output is defined based on the Taylor distribution for a sidelobe level (SLL) of $-20 \ \mathrm{dB}$ \cite{Balanis:AntennaBook}, specifically
\begin{align} \label{eq:E_taylor}
E^{(2)}_{\mathrm{out},z}(x) = A J_0\left[j\pi B \sqrt{1-\left(\frac{2 x}{l}\right)^2}\right],
\end{align}
where $J_0$ is the Bessel function of the first kind and zero order, $l=6 \lambda$ is the total length of the Taylor aperture, $B=0.7386$ is a parameter that sets the SLL to the desired level of $-20 \ \mathrm{dB}$ and $A$ is a normalization parameter to equalize the total output power at the upper side with the total incident power, as defined in Eq.~\eqref{eq:Pav2}. The magnetic field is computed using the plane-wave decomposition, as described for the input and output distributions in the previous example.

\begin{figure}
\centering
{\includegraphics[width=0.8\columnwidth, trim=0 0 0 0 0, clip=true]{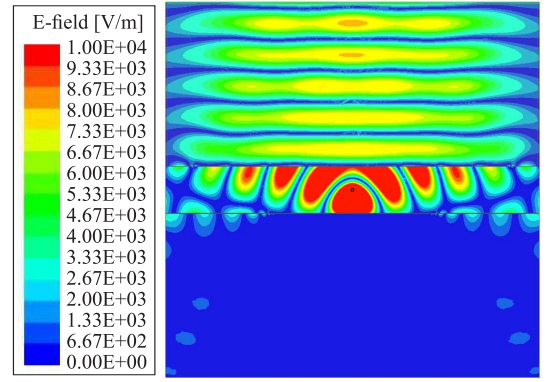}}
\caption{Real part (absolute values) of the electric field $E_z$ for the single-side Taylor antenna with a sidelobe level of $-20 \ \mathrm{dB}$. The source is located in the middle of the two metasurfaces that are placed $d=0.75 \lambda$ apart.}
\label{fig:Fig9}
\end{figure}

\begin{figure}
\centering
{\includegraphics[width=0.9\columnwidth, trim=0 0 0 0 0, clip=true]{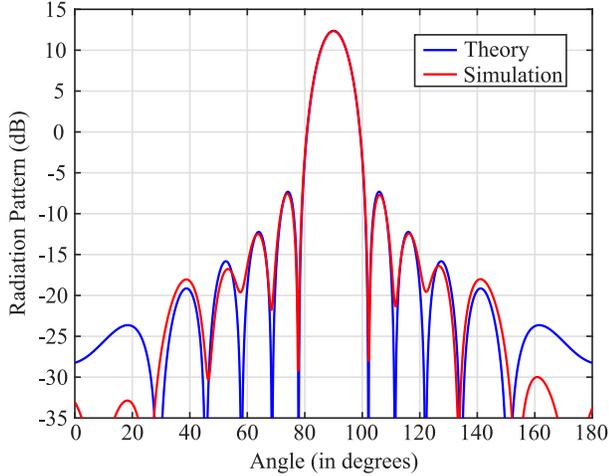}}
\caption{Radiation pattern for the single-side designed Taylor antenna. The simulation results (red line) compare well with the theoretical Taylor pattern (blue line) with the directivity, the sidelobe level and the HPBW being all very close to the desired values.}
\label{fig:Fig10}
\end{figure}

Having determined the incident and the output fields at the two metasurfaces, the proposed optimization method is used to design the metasurface pair and the total structure is simulated in ANSYS HFSS. The real part of the electric field $E_z$ is plotted in Fig.~\ref{fig:Fig9}, where it is clear that the waves destructively interfere in the lower output resulting in very low transmission to this side, as desired. In addition, most of the power provided from the source escapes from the upper output according to the amplitude and phase profile defined in Eq.~\eqref{eq:E_taylor}. 
To verify the accuracy of the transformation, the far-field radiation pattern is calculated and it is compared with the theoretical one, as predicted from the equivalence principle applied at the output of M2. As it is observed in Fig.~\ref{fig:Fig10}, the curve obtained from the simulation matches very closely with the theoretical one regarding the maximum directivity, the amplitude of each lobe and the locations of the nulls. In particular, the simulated SLL and HPBW are $-19.91 \ \mathrm{dB}$ and $9.02 \degree$ compared to theoretical values of $-19.7 \ \mathrm{dB}$ and $8.94 \degree$, respectively. It should be noted that the theoretically predicted SLL is not exactly $-20 \ \mathrm{dB}$, because the equivalence principle takes into account the discretization of the output fields, according to the width of the unit cells. More significant deviations between theory and simulation results can be observed in Fig.~\ref{fig:Fig10} for angles far away from broadside. However, the amplitude in these angles ($\phi < 45 \degree$ and $\phi > 45 \degree$) is more than $30 \ \mathrm{dB}$ lower compared to broadside; thus, even a small power-density mismatch at the boundaries or numerical issues can affect the simulated values. Regarding the power efficiency, it is calculated that $95.4 \%$ of the total input power is transmitted to the upper output, while the power leakage below M1 and to the sides is only $0.7 \%$ and $3.9 \%$, respectively. It is pointed out that the total power at the upper output slightly exceeds the total incident power (accounting for $92.1$ of the input power), as defined in Eq.~\eqref{eq:Pav2}. This is attributed to the partial cancellation of the power leaking to the sides, as the cylindrical wave that directly escapes the metasurface pair interferes destructively with the power escaping due to reflections at the inner boundaries of the two metasurfaces.

\subsection{Double-sided beamforming with a single line-source in the middle of the metasurface pair}
The beamforming example shown in the previous section can be extended so that both output sides support two independent field distributions. This case can be perceived as a generalization of the cavity-based antenna presented in \cite{Epstein:TAP2017Antennas}, where the PEC wall used there is replaced with a second omega-bianisotropic HMS. For this scenario, the geometry is identical to the one in Sec.~\ref{sec:Single-sided Taylor}, but the aim is to have a Taylor-antenna output at both sides of the metasurface pair instead of redirecting all the incident power to the upper side. The Taylor antenna characteristics at both sides are the same as in Sec.~\ref{sec:Single-sided Taylor} (SLL of $-20 \ \mathrm{dB}$ and aperture length $l=6 \lambda$) with the additional requirement that the lower-side Taylor antenna has its maximum directivity at the azimuthal angle $\phi=-110 \degree$ ($20 \degree$ off-broadside). Therefore, the output electric field at M1 is not set to zero, but takes the form of Eq.~\eqref{eq:E_taylor} with an additional linear phase to account for the tilting of the maximum directivity direction. Lastly, the amplitudes of the field distributions are normalized so that the total incident power is equally divided at the two sides. It is emphasized that any other unequal power splitting would also be possible, as long as it sums up to the total incident power defined in Eq.~\eqref{eq:Pav2}.

\begin{figure}
\centering
{\includegraphics[width=0.8\columnwidth, trim=0 0 0 0 0, clip=true]{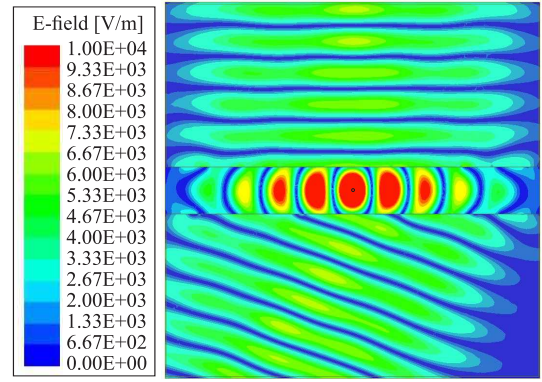}}
\caption{Real part (absolute values) of the electric field $E_z$ for the double-sided output Taylor antenna with a sidelobe level of $-20 \ \mathrm{dB}$ for both sides and a tilt of $20 \degree$ off-broadside for the lower output side. The source is located in the middle of the two metasurfaces that are placed $d=0.75 \lambda$ apart.}
\label{fig:Fig11}
\end{figure}

The real part of the electric field, as given by full-wave simulations, is depicted (in absolute values) in Fig.~\ref{fig:Fig11}. It is clear that the two metasurfaces form a cavity and the power leaks from both output sides according to the specified field distributions. In particular, the output aperture at M2 radiates towards broadside ($\phi=90 \degree$), while the output aperture at M1 radiates $20 \degree$ off-broadside ($\phi=-110 \degree$). As expected, the field values are higher within the metasurface pair, where the multiple reflections produce a higher power concentration. However, as we move horizontally away from the center, the values continuously decay and only a relatively small portion of the power leaks from the two open sides.

\begin{figure}
\centering
{\includegraphics[width=\columnwidth, trim=0 0 0 0 0, clip=true]{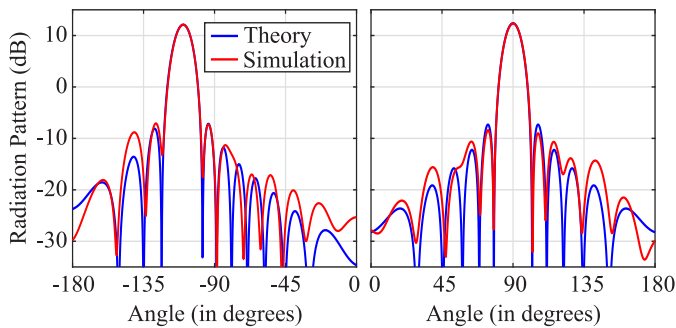}}
\caption{Radiation pattern for the lower side (left) and upper side (right) of the double-sided Taylor antenna. For both sides, the simulations (red curves) match satisfactorily with the theoretical patterns (blue curves), especially close to the angles of maximum radiation.}
\label{fig:Fig12}
\end{figure}

The radiation pattern is calculated independently for the two output sides, by selecting the respective boundaries to perform the near-field to far-field transformation. The simulated results are given in Fig.~\ref{fig:Fig12} and they compare well with the theoretical radiation patterns obtained from the equivalence principle. It is clear that the power is radiated at both sides towards the desired angles. The difference between simulation and theory regarding the directivity and the HPBW is less than $0.05 \ \mathrm{dB}$ and $0.15 \degree$, respectively, for both sides. Moreover, the sidelobe level is only $1 \ \mathrm{dB}$ higher at the lower output and $1.1 \ \mathrm{dB}$ lower at the upper output compared to the predicted values. While some deviations from theory exist for the minor lobes at both outputs, it can be seen that they generally decay away from the angle of maximum radiation, as expected. Lastly, regarding the power splitting, it is estimated that $50.9 \%$ and $48.5 \%$ of the total input power is guided to the lower and upper output, respectively, while only $0.6 \%$ escapes from the two sides.
 
\section{Cavity effects on the sensitivity of the metasurface pair} \label{Sec:CavityEffects}
From the previous examples, it can be observed that the introduction of multiple reflections into the design leads to an energy build-up within the metasurface pair. As demonstrated, this cavity effect is essential to reduce the separation distance between the two HMSs, when designing for wave transformations that do not locally conserve power. However, the internal reflections induced by the two partially reflective HMSs can substantially decrease the bandwidth of the proposed structure and, in general, affect its sensitivity with respect to the geometrical parameters involved. This effect can be interpreted as an outcome of the interaction between the two metasurfaces, as small errors due to variations in the frequency or the geometry will accumulate as the propagating wave reflects multiple times at the two boundaries. Alternatively, perceiving the structure as a cavity, the multiple reflections increase the energy stored within the cavity with respect to the total radiated power, thus resulting in a higher quality factor and a smaller bandwidth. 

In this section, the trade off between the separation distance and the sensitivity of the structure is investigated, based on the example of the uniform output aperture presented in Sec.~\ref{Sec:UniformOutput}. Since a bandwidth analysis would require the dispersion of each unit cell to be taken into account, we choose to study the sensitivity of the structure instead, by varying the distance $d$ between the two metasurfaces in a range around its nominal value. In this way, the parameters of the two metasurfaces remain constant and it is possible to study separately only the effect of multiple reflections in the sensitivity of the structure. Moreover, some intuition into the expected frequency bandwidth can be obtained (disregarding the dispersion of the HMSs), as a change in frequency can be translated into a change of the electrical distance between the two metasurfaces.

\begin{figure}
\centering
{\includegraphics[width=0.9\columnwidth, trim=0 0 0 0 0, clip=true]{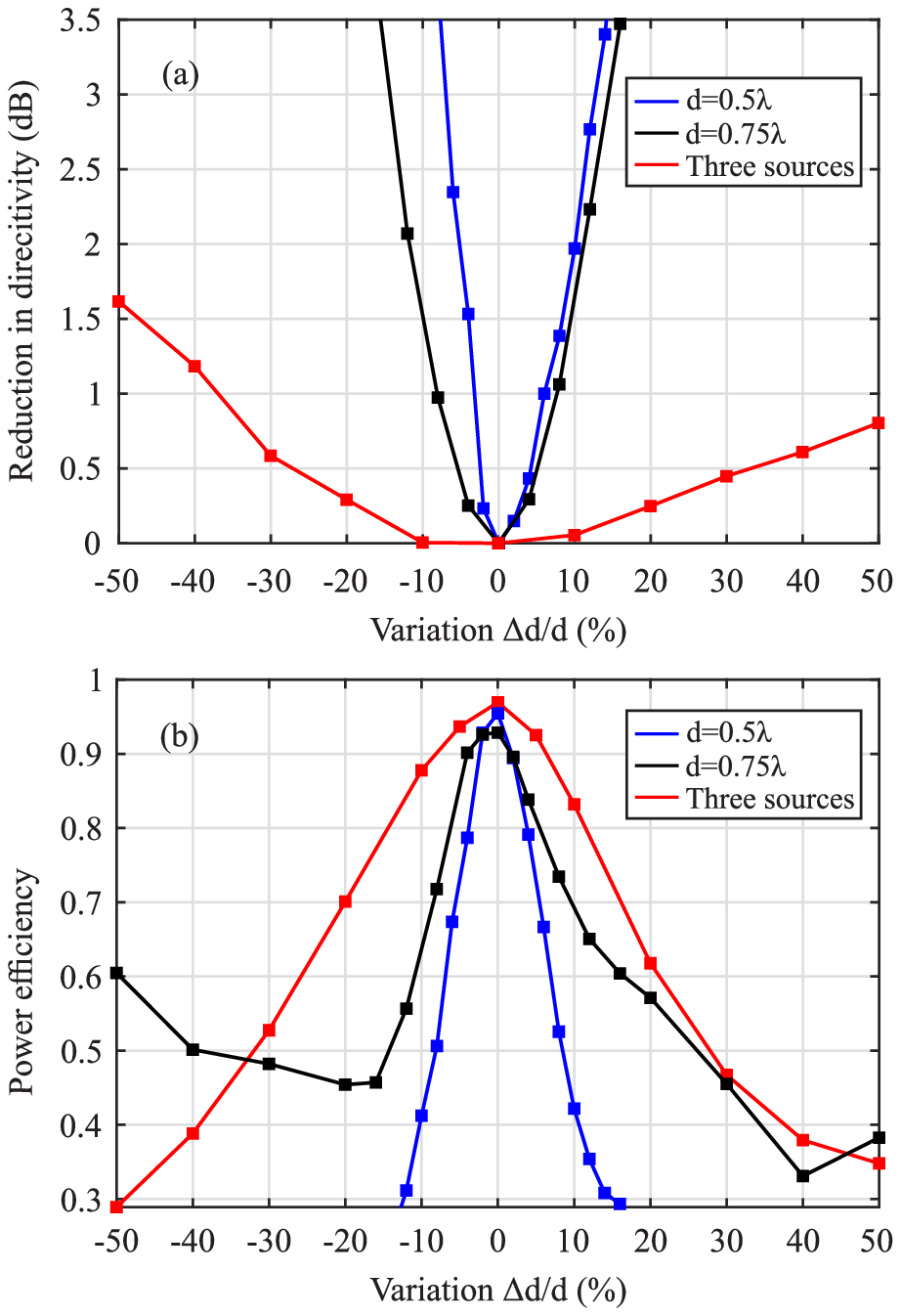}}
\caption{Parametric analysis with respect to the distance $d$ between the two metasurfaces for the example presented in Sec.~\ref{Sec:UniformOutput}. The sensitivity of both the directivity (top figure) and the power efficiency (bottom figure) can be improved by increasing the distance to $d=0.75\lambda$ (black curves) or by illuminating better M1 using three sources (red curves). The marks represent the cases simulated to obtain the curves for each design.}
\label{fig:Fig13}
\end{figure}

First, we vary the distance $d$ for the example including multiple reflections, as designed in Sec.~\ref{Sec:UniformOutput} for a nominal distance $d=0.5 \lambda$. The reduction in the maximum directivity and the power efficiency, defined as the ratio between the total output power and the total incident power at M1, are calculated for each variation and the results are plotted in Fig.~\ref{fig:Fig13}. It is clear that the structure is highly sensitive, since only a $\pm 4 \%$ variation in $d$ results in approximately $1~\mathrm{dB}$ drop of the directivity and noticeable reflections, as the power efficiency drops to around $80 \%$. The performance becomes even worse for larger variations of the distance $d$, implying that an implementation of such a structure would be challenging, especially for higher frequencies. To compensate for the cavity effects, depending on the application, a slightly larger distance may be preferred for certain applications. The wave transformation is redesigned using multiple reflections for a nominal distance $d=0.75 \lambda$. While the radiation pattern for the designed distance is very close to the ideal, the sensitivity with the relative distance variation $\Delta d/d$ is milder, as confirmed from Fig.~\ref{fig:Fig13}. Regarding the power efficiency for this case in Fig.~\ref{fig:Fig13}(b), we note that results for relatively large variations of the distance ($|\Delta d/d| >0.2$) do not provide any useful information, as a significant portion of the output power is transmitted to beams away from broadside. 

Finally, another way to mitigate the cavity effects without changing the separation distance $d$ would be to modify the input power density so that it is more distributed along M1. To explore this possibility, two current line-sources are added at the same distance $s=\lambda/3$ from M1, but displaced by $\pm 1.25 \lambda$ in the $x$-direction with respect to the center source. In addition, the current amplitude of the edge sources is set to be three times less than the current amplitude of the center one. The optimization procedure is slightly modified, as the structure is first designed with purely transmissive omega-bianisotropic HMS ($A_{r,n}=0$). The solution is then used as a starting point for the optimization algorithm including multiple reflections. The outcome of this two-step optimization process is to keep the amplitude of the backward-propagating wave at a relatively low value compared to the forward-propagating wave. The wider illumination of M1 combined with the modification of the optimization process leads to greater immunity of the transformation performance when varying the distance $d$ away from its nominal value, as is evident for this case in Fig.~\ref{fig:Fig13}. In particular, it can be observed that for a  $\pm 20 \%$ variation of the separation distance, the directivity drops less that $0.3 \ \mathrm{dB}$, while the power efficiency remains above $60 \%$. Practically, it is expected that the frequency bandwidth in this case would be limited from the dispersion of the individual metasurfaces, as dictated by the unit cell design, rather than the sensitivity to separation distance $d$. While the structure is no longer single-fed and three sources should be controlled independently, this choice can also be considered, if it is necessary to maintain compactness and design a metasurface pair that is less prone to geometrical or frequency variations.

\section{Conclusion}  \label{Sec:Conclusion}
In conclusion, wave transformations that do not locally conserve real power have been designed using pairs of Huygens' metasurfaces. A design method has been developed, based on determining the electric and magnetic fields between the two metasurfaces, so that the power density is matched simultaneously at the two boundaries. The method relies on expanding the field distributions as weighted basis-functions summations and, then, minimizing the power mismatch across the metasurfaces through optimization of the unknown weights. By allowing for multiple reflections, it was shown that the required distance between the two metasurfaces can be significantly reduced, resulting in compact, yet accurate, field transformations. The design method is also adjustable to applications which require the source to be located within the metasurface pair. Through the example of a Taylor-pattern antenna fed by a single line-source, it was shown that a desired output can be supported at one or both sides of the metasurface pair. Finally, the sensitivity with respect to the separation distance between the two metasurfaces was discussed, when multiple reflections are present, revealing a trade off between the bandwidth and the size of the structure.

\section*{Acknowledgement}
Financial support from the Natural Sciences and Engineering Research Council of Canada is gratefully acknowledged. V.~A. also acknowledges the support of the Ministry of Advanced Education and Skills Development of Ontario, the Electrical and Computer Engineering Department of the Univ. of Toronto and the Alexander S. Onassis Foundation.

The authors thank Michael Chen for helpful discussions.

\bibliographystyle{ieeetran}

\end{document}